\documentclass[twocolumn,trackchanges]{aastex631}

\usepackage{tikz}

\newcommand{\affcam}{DAMTP, Centre for Mathematical Sciences, University of Cambridge, Wilberforce Road, Cambridge CB3 0WA, UK}
\newcommand{\affcamast}{Institute of Astronomy, University of Cambridge, Madingley Road, Cambridge, CB3 0HA, UK}

\newcommand{\affckavli}{Kavli Institute of Cosmology (KICC), University of Cambridge, Madingley Road, Cambridge, CB3 0HA, UK}

\begin{document}

\title{\large Constraining Dark Energy from Local Group dynamics}

\author[0000-0002-9578-3081]{David Benisty}
\affiliation{\affcam}\affiliation{\affckavli}
\author{Anne-Christine Davis}
\affiliation{\affcam}\affiliation{\affckavli}
\author[0000-0002-5981-7360]{N. Wyn Evans}
\affiliation{\affckavli}\affiliation{\affcamast}
\correspondingauthor{nwe22@cam.ac.uk}

\begin{abstract}
This {\it Letter} develops a method to constrain the Cosmological Constant $\Lambda$ from binary galaxies, focusing on the Milky Way and Andromeda. We provide an analytical solution to the two-body problem with $\Lambda$ and show that the ratio between the Keplerian period and $T_\Lambda = 2\pi/(c \sqrt{\Lambda}) \approx 63.2$ Gyr controls the importance of effects from the Cosmological Constant. The Andromeda-Milky Way orbit has a period of $\sim 20$ Gyr and so Dark Energy has to be taken into account.  Using the current best mass estimates of the Milky Way and Andromeda galaxies, we find the Cosmological Constant value based only on the Local Group dynamics to be lower then $5.44$ times the value obtained by Planck. With future astrometric measurements, the bound on the Cosmological Constant can be reduced to $\left(1.67 \pm 0.79\right) \Lambda_{\rm PL}$. Our results offer the prospects of constraints on $\Lambda$ over very different scales than previously. The Local Group provides also a completely novel platform to test alternative theories of gravity. We illustrate this by deriving bounds on scalar-tensor theories of gravity over Megaparsec scales. 
\end{abstract}


\section{Introduction}
The discovery that the cosmic expansion is accelerating puzzled scientists and led them to propose the existence of a mysterious \lq\lq dark energy" that makes up 68\% of the energy of the observable universe \citep{Perlmutter:1998np,Riess:2019cxk,Planck:2018vyg}. The Cosmological Constant is also important on the local Universe scale \citep[e.g.,][]{Chernin:2001nu,Kim:2020gai,Karachentsev:2003eh,Silbergleit:2019oyx}. In particular, the Cosmological Constant changes the predicted mass of the Local Group~\citep[e.g.,][]{Pa13, Penarrubia:2014oda,Hartl:2021aio,Benisty:2019fzt,Be22}. In this {\it Letter}, we give a bound on $\Lambda$ from Local Group dynamics which is compatible with the cosmological measurements.  {The novelty here is that this provides an exploration of dark energy on very different scales.} We extend this idea to other Dark Energy models and show that this  {method can also constrain modified gravity theories.}

\section{Analytical Solution of the Two Body Problem with $\Lambda$}
\label{sec:form}

The dynamics of two bodies that include the Cosmological Constant contribution is governed by the action \citep{Carrera:2006im}:
\begin{equation}
\mathcal{L} = \frac{1}{2} v^2 - \frac{G M}{r} - \frac{1}{6} \Lambda c^2 r^2, 
\label{eq:Lag}
\end{equation}
where $r$ is their mutual separation, $v$ is the total velocity, $\Lambda$ is the Cosmological Constant, $c$ is the speed of light and $G$ is the Newtonian Constant. We use $\mu$ as the reduced mass and $M$ as the total mass of the system. The conservation of energy and  angular momentum reads
\begin{equation}
\epsilon = \frac{1}{2}\dot{r}^2 +\frac{l^2}{2 r^2} -\frac{G M}{r} + \frac{1}{6} \Lambda c^2  r^2
,\quad l = r^2 \dot{\theta},
\label{eq:enerLam}
\end{equation}
where $\epsilon$ is the total energy per reduced mass, $l$ is the total angular momentum per reduced mass, while a dot denotes the derivative with respect to time and $\theta$ is the true anomaly. In order to find the relation between the energy and the angular momentum to the eccentricity and the semi-major axis, we use the boundary condition that the time derivative of the separation is zero at the extremal separations (pericentre and apocentre) $\dot{r}\left[a\left(1 \pm e\right)\right] = 0$. Inserting this condition into Eq.~(\ref{eq:enerLam}) yields the relations:
\begin{equation}
\epsilon = -\frac{G M }{2a} \left(1-\frac{2}{3} \lambda \left(1 + e^2\right) \right)
\end{equation}
\begin{equation}
l^2 =  G M a \left(1 - e^2\right) \left(1 + \frac{\lambda}{3}\left( 1 - e^2\right) \right)
\end{equation}
where we have introduced the dimensionless parameter:
\begin{equation}
\lambda := \frac{a^3 c^2}{G M}\Lambda = \left(\frac{T_{\text{Kep}}}{T_{\Lambda}}\right)^2,
\end{equation}
that quantifies the impact of the Cosmological Constant. Here, $T_{\text{kep}}$ is the Keplerian period of the system and $T_{\Lambda}$ is a period that relates the cosmological constant:
\begin{equation}
T_{\text{Kep}} = 2 \pi \sqrt{\frac{a^3}{G M}}, \quad  T_{\Lambda} = \frac{2 \pi }{c \sqrt{\Lambda}}  = \left(63.22 \pm 0.16\right) {\rm Gyr}.  
\end{equation}
$T_{\Lambda}$ is calculated from the measured Planck values~\citep{Aghanim:2018eyx} and is much larger than the age of the Universe $\sim 13.8\, $Gyr. In order to solve the motion, it is useful to parameterize the separation as
\begin{equation}
r/a = 1 - e \cos \eta,
 \label{eq:meanAnomDef}
\end{equation}
where $e$ is the eccentricity, $a$ is the semi-major axis and $\eta$ is the eccentric anomaly. Putting this parameterizations into~(\ref{eq:enerLam}) gives a differential equation (that relates $\dot{\eta}$ into $\eta$). The limit $\lambda < 1$ is readily tackled with perturbation theory. Up to the first order correction, we obtain:
%
%
%
\begin{equation}
\frac{\sqrt{G M /a^3} }{1-e c} \frac{d t}{d\eta} \approx  1 - \frac{\lambda }{12}  \left( 8 - e (8 c_1 -c_2 e+e)\right),
\end{equation}
where $c_1 = \cos \eta$ and $c_2 = \cos 2 \eta$ and etc. Integration gives the solution:
\begin{eqnarray}
&&
\frac{2\pi}{T}\left(t - t_0\right) = \eta - e_t \sin \eta  -\lambda \left(\frac{5e^2}{24}   s_2 - \frac{e^3}{72} s_3\right).
\label{eq:motiontime}
\end{eqnarray}
where $s_1 = \sin \eta$ and $s_2 = \sin 2 \eta$ etc. with the couplings:
\begin{equation}
\frac{e_t}{e} = 1 - \frac{16 - 7 e^2}{24} \lambda, 
\quad \frac{1}{T} = \frac{1}{T_{\text{Kep}}} \left[1 - \frac{8 + 3 e^2}{12} \lambda\right].    
\end{equation}
In order to find the relationship between the true and the eccentric anomaly, we use the conserved angular momentum from Eq.~\ref{eq:enerLam}. Integrating the derivative of the true anomaly with respect to the eccentric anomaly gives:
%
%
%
\begin{equation}
2\pi\frac{\theta - \theta_0}{\Phi} = \tilde{\nu}_e - \frac{\lambda}{6} \sqrt{1-e^2}  \left(3 \eta  - e \sin \eta \right),
\label{eq:motiontheta}
\end{equation}
where
\begin{equation}
\tan\frac{\tilde{\nu}_e}{2} \equiv \sqrt{\frac{1+e}{1-e}} \, \tan \frac{\eta}{2}, \quad \Delta\theta = \frac{\Phi}{2 \pi} - 1 = \frac{\lambda}{2} \sqrt{1-e^2},
\end{equation}
where $\theta_0$ is the initial true anomaly, $\Delta \Phi$ is the contribution of the precession from $\Lambda$ which coincides with earlier expressions~\citep{Kerr:2003bp,Iorio:2007ub}. The two equations~(\ref{eq:motiontime}) and~(\ref{eq:motiontheta}) describe a precessing elliptical orbit, which is the full motion of the binary system up to first order in $\lambda$. 

Besides the orbital equations, the tangential and the radial velocity terms are useful, since these are related to the velocities we measure. Using $v_r = \dot{r}$ and $v_t = \dot{\theta} r =l^2/r^3$ relations, we get:
\begin{equation}
v_r^2 = \frac{G M}{a}\left(\frac{e s}{1 - e c}\right)^2 \left[1 + \frac{\lambda}{6}   \left(8 + e^2 c_2-e^2-8 e c \right)\right],
\label{eq:vr}
\end{equation}
\begin{equation}
v_t^2 = \frac{G M}{a} \frac{1-e^2}{\left(1- e c\right)^2} \left(1 +\frac{1-e^2}{3}\lambda \right),
\label{eq:vt}
\end{equation}
%

The Timing Argument (TA)~\citep{Ka59} for the Local Group assumes that the Milky Way (MW) and Andromeda (M31), modelled as point masses, have been approaching each other despite cosmic expansion. Briefly after their formation (for the sake of simplicity the \lq\lq Big Bang"), these two galaxies must have been in the same place with zero separation ($r=0$). Due to the Hubble expansion, these two galaxies moved apart. After a couple of billion years, they slowed down and then moved towards each other again as a consequence of the gravitational pull. Many authors~\citep{Penarrubia:2014oda,Chamberlain:2022fqr,Sawala:2022ayk} consider the dynamics as two body motion without the Cosmological Constant. Here, for the first time, we have modified these terms analytically in order to track the dynamics in the presence of $\Lambda$. Using values for the Local Group taken from \citet{Be22}, based on the proper motion measurements of \citet{vanderMarel:2012xp} and \citet{Sa21}, we find $\lambda = 0.103 \pm 0.002$. Since $\lambda < 1 $, the analytical approximation is good.


Given the simplicity of the TA, it is now customary to include corrections motivated from the galaxy formation simulations to get unbiased estimates of the true Local Group mass~\citep[e.g.,][]{Li:2007eg}. The matter has been re-investigated recently by \cite{Hartl:2021aio}, who used N-body and hydrodynamical simulations. The tendency of the TA is to overestimate the Local Group mass by a factor of $0.63 \pm 0.02$~\citep{Be22}. 

\begin{figure}[t!]
    \centering
\includegraphics[width=0.48\textwidth]{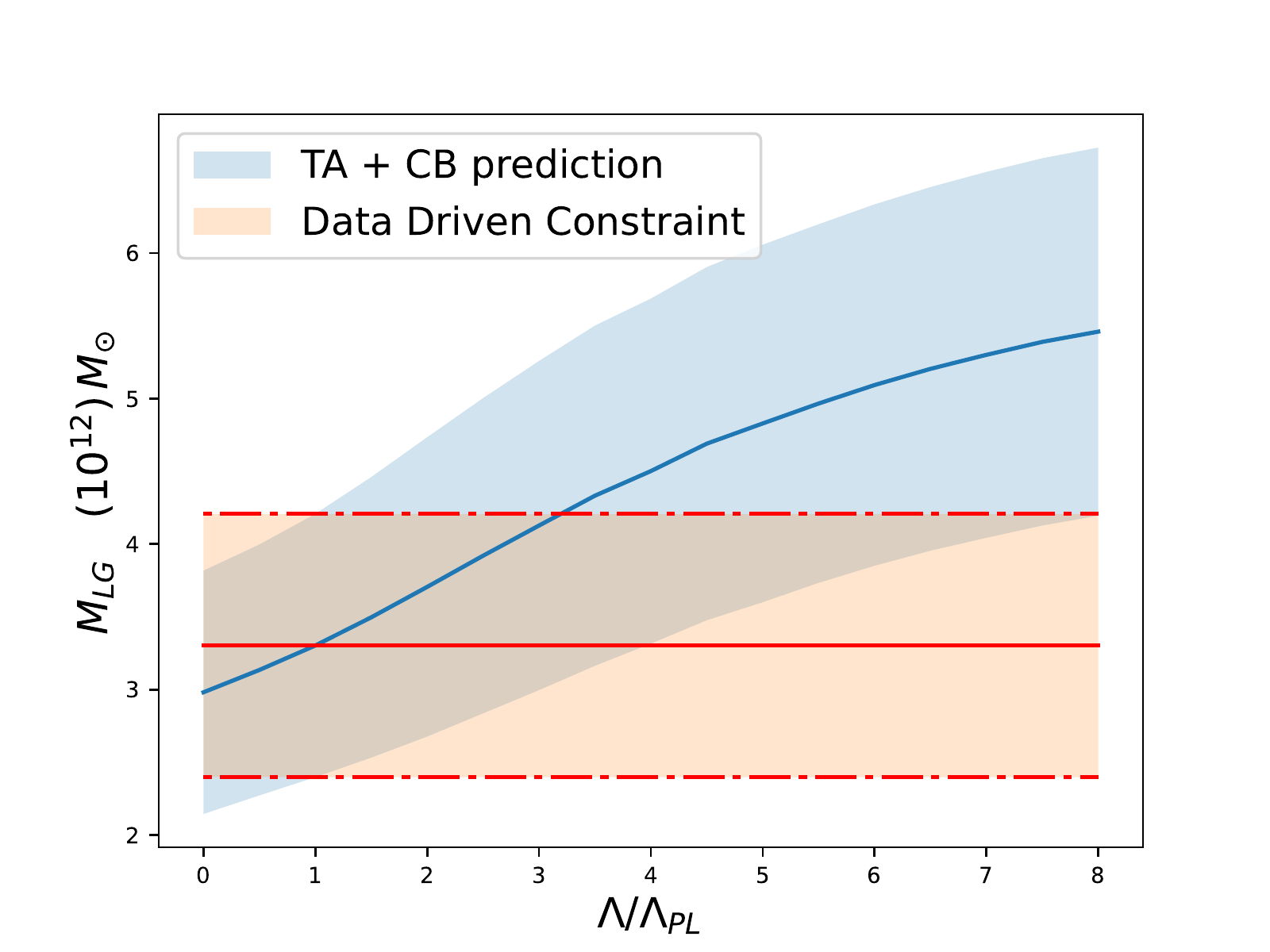}
    \caption{\it{The Timing-Argument + Cosmic Bias prediction for the Local Group mass vs. the Cosmological Constant $\Lambda$ value (blue). The red curve shows the latest updated mass of $(3.2 \pm 1.6) \, 10^{12} M_\odot$ from \cite{Be22}. }}
    \label{fig:effpe}
\end{figure}

\section{Bounds on dark energy and modified gravity}

The Local Group mass depends on dark energy. By adding up the mass associated with the galaxies in the Local Group, \citet{Be22} estimated it is $\left(3.7 \pm 0.5\right) \times 10^{12} M_\odot$ \cite[see also ][for a very similar value]{Sawala:2022ayk}. From this data-driven measurement, we can constrain the Cosmological Constant, as well as modifications of the Newtonian force on the scales of Mpcs.

\begin{figure}
    \centering
\includegraphics[width=0.42\textwidth]{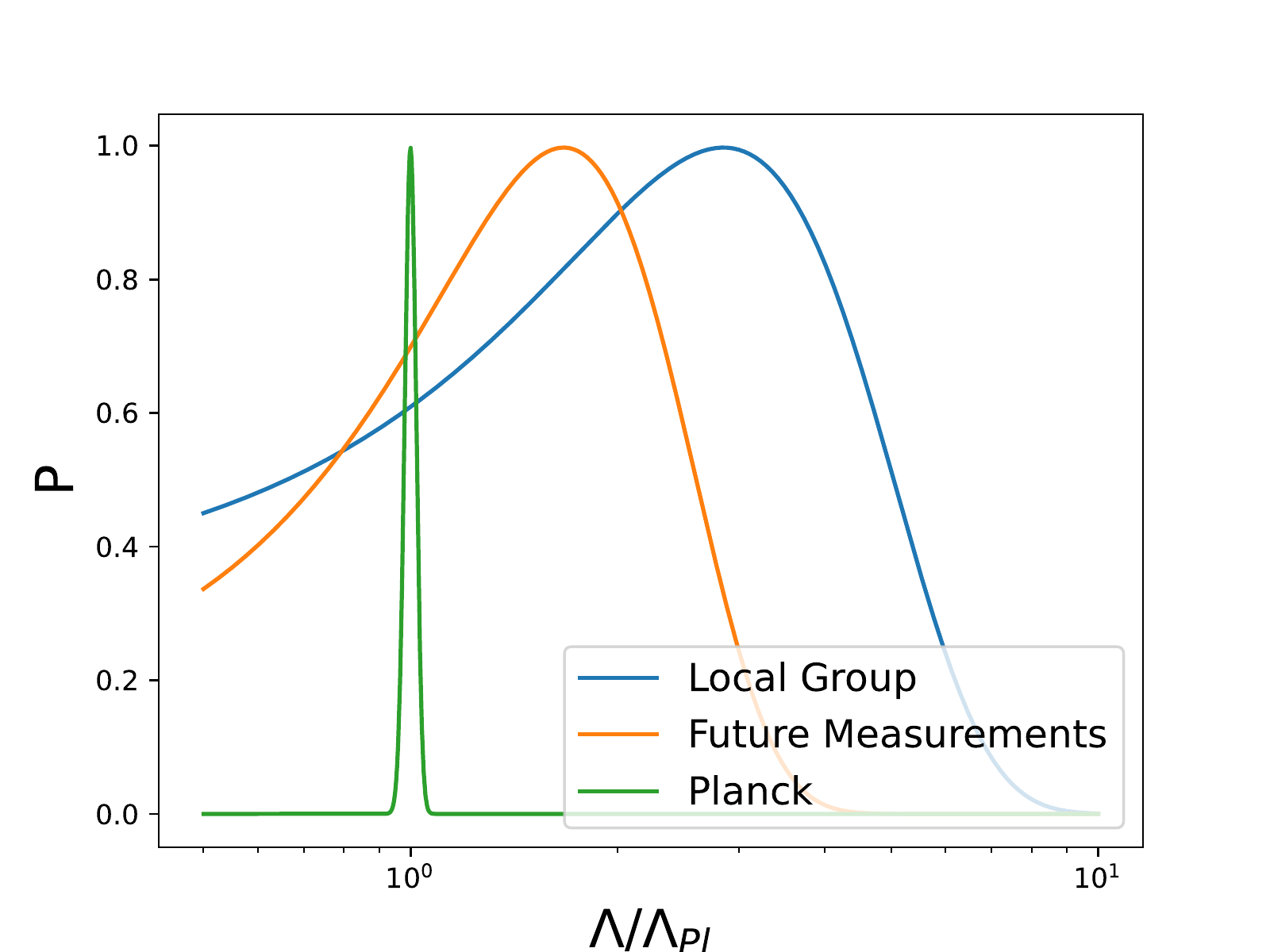}
\\
\includegraphics[width=0.42\textwidth]{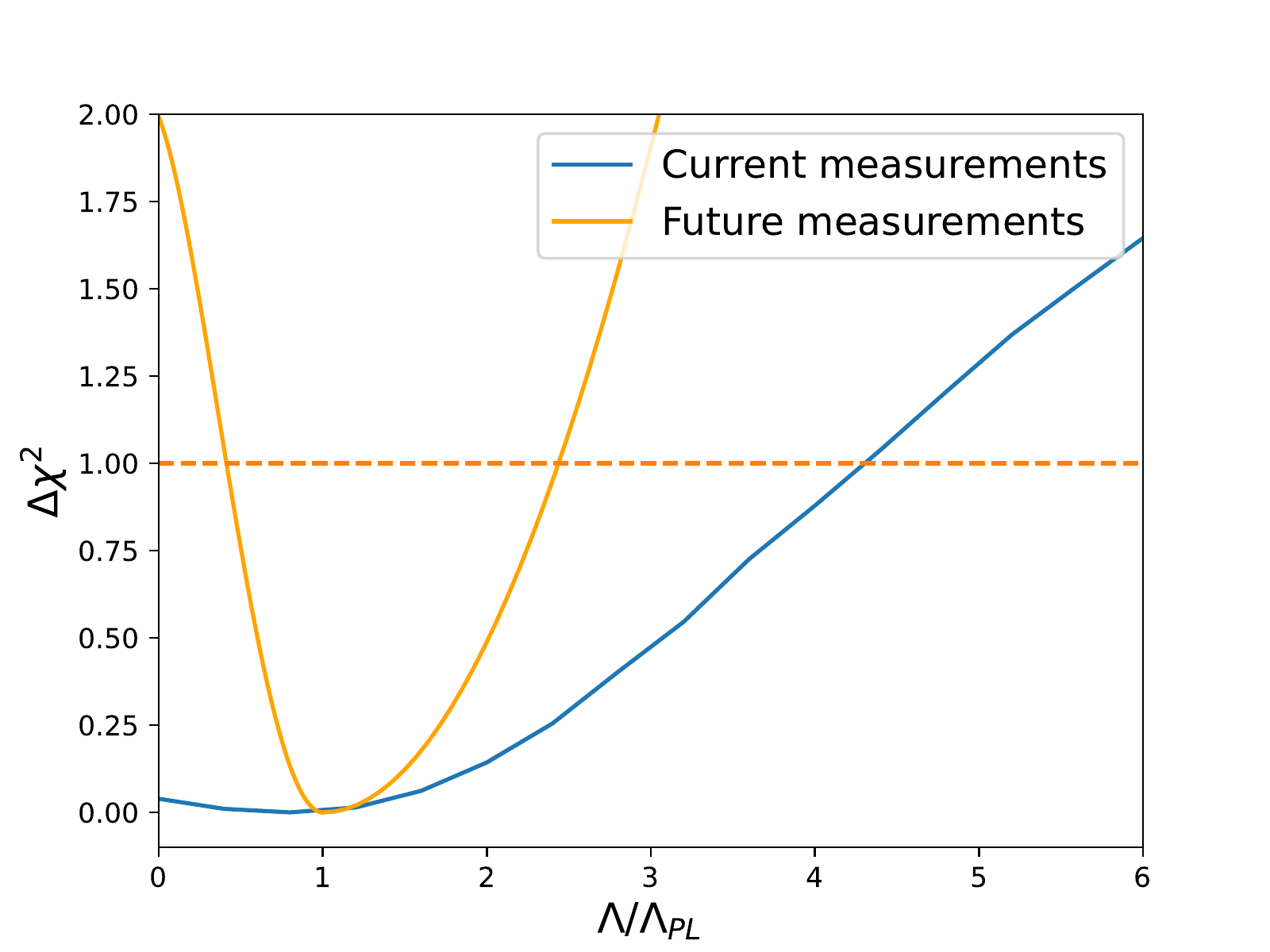}
\caption{\it{  {Top: Bound on $\Lambda$ from the Local Group dynamics (blue) compared to the measured value from Planck (green). The special case (orange) is for $5\%$ for the mass and for the tangential velocity, with $2\%$ for the other quantities. For this case, the upper bound on $\Lambda$ is predicted to be $1.67\pm 0.79$ times from the Planck value. Bottom: The reduced $\Delta \chi^2$ for the current and the future LG measurements.} }}
    \label{fig:postLambda}
\end{figure}

\subsection{Cosmological Constant}
 {Fig.~\ref{fig:effpe} shows the Local Group mass for different values of the Cosmological Constant, normalised to the Planck value. Drawing from the distributions of separations $r$, radial velocities $v_r$ and tangential velocities $v_t$ provided in \citet{Be22}, we show the posterior distribution for $M$ at each $\Lambda$. The horizontal red curves show the latest observationally derived mass and its upper and lower bounds. Requiring consistency, we define the reduced $\chi^2$ as the difference with respect to the data driven mass:}
\begin{equation}
\Delta \chi^2 =  \left(\frac{M (r,v_r, v_t, \Lambda) - \bar{M}_{\rm LG}}{\Delta M_{\rm LG}} \right)^2,
\end{equation}
 {where $M (r,v_r, v_t, \Lambda)$ is the predicted mass for different values of $\Lambda$ and $\bar{M}_{\rm LG} \pm \Delta M_{\rm LG}$ is the data driven one. The lower panel of Fig.~\ref{fig:postLambda} shows the $\Delta \chi^2 $ for different values of $\Lambda$. Using a broad prior of $\Lambda/\Lambda_{\rm PL} = [0,10]$, we obtain the posterior on $\Lambda$ to be:}
\begin{equation}
 {\Lambda/\Lambda_{\rm PL} = 3.074 \pm 2.369.}
\end{equation}
 {The upper panel of Fig.~\ref{fig:postLambda} compares the bound on $\Lambda$ from the Local Group dynamics with the measured value from Planck. The value of Planck sits with $1\sigma$ of the distribution, and the upper bound is $\preceq 5.44 \, \Lambda_{\rm PL}$. Since the $\Delta \chi^2$ is flattish, with the broad uniform prior we get weak constraints on the corresponding $\Lambda$ value.}

 {This value of $\Lambda$ will greatly improve in the near future thanks to new data. In the lower panel, we now examine the effects of (i) improved proper motions, (ii) reduced cosmic bias uncertainty, (iii) better observational constraints on the mass of the Local Group. From the data summary provided in \citet{Be22}, the M31-MW separation has an uncertainty of $\delta r = 5.2 \%$, whilst the velocities have uncertainties $\delta v_r = 4 \%$, $\delta v_t = 37.86  \%$. However, improved proper motions for M31 are likely to come quickly thanks to the ability of {\it JWST} to obtain high precision astrometry in the Local Group, often exceeding the accuracy of {\it Gaia}~\citep{Li23}. The main uncertainty in the mass of the Local Group is the contribution of its largest member, M31. This is inferred from the kinematics of its satellite galaxies and so will also benefit from {\it JWST} capabilities. In orange we show an optimistic case, taking the mass and tangential velocity to have a  $5\%$ error and the other quantities to have a $2\%$ error. This gives the value to be:}
\begin{equation}
     {\Lambda/\Lambda_{\rm PL} = 1.670 \pm 0.794}.
\end{equation}

 {From the posterior, we predict that it will be possible not only to give strong upper bound on $\Lambda$ which is in order of the Planck value, but will give a detection of $\Lambda$ distribution with $\sim 40\%$ error.}

\begin{figure}
    \centering
\includegraphics[width=0.45\textwidth]{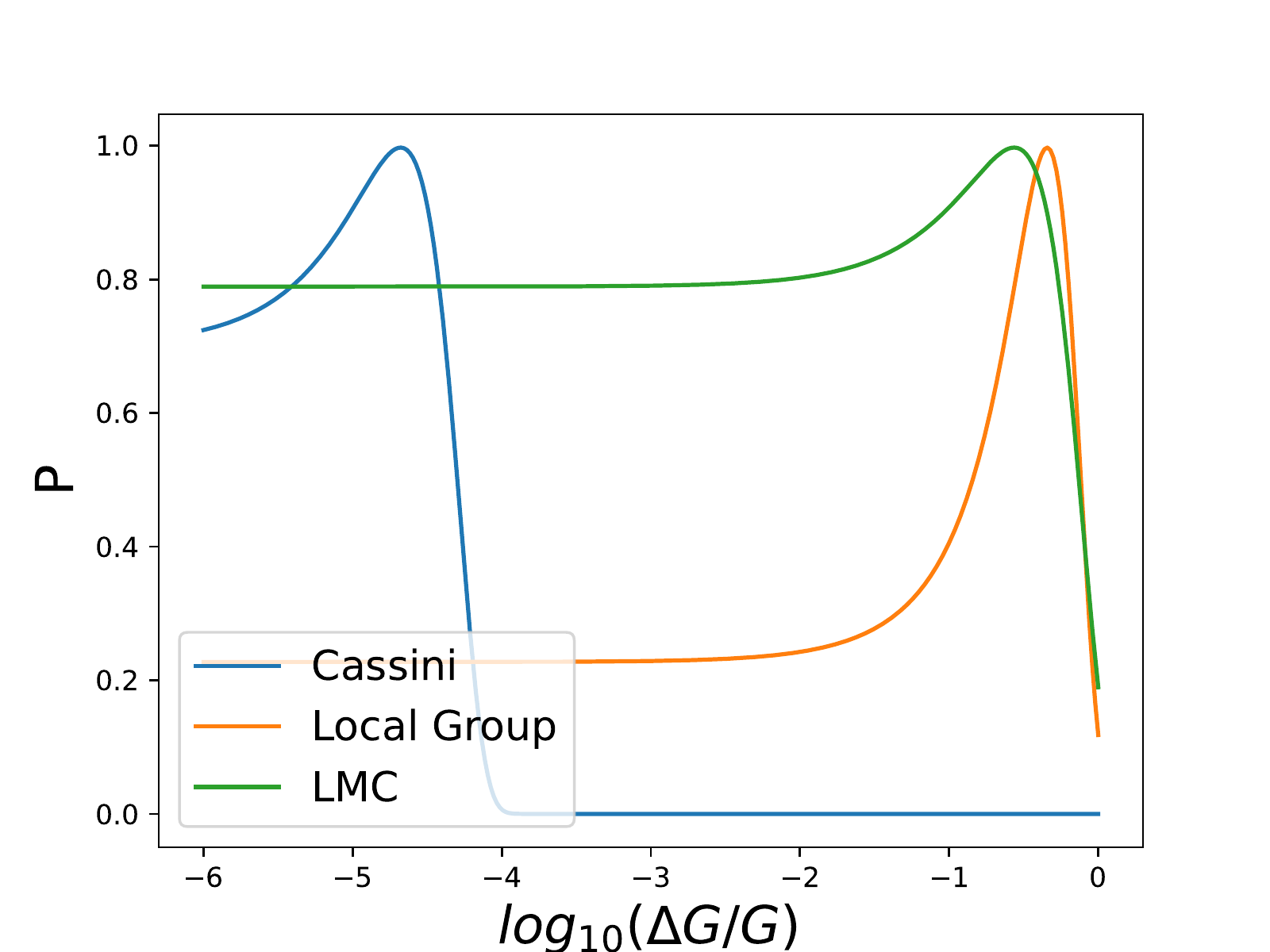}
\caption{\it{Bound on the running Gravitational Constant $\Delta G/G$ from the Local Group dynamics versus the Cassini bound~\citep{Bertotti:2003rm} and Cepheids in the Large Magellanic Cloud ~\citep{Desmond:2020nde}}}
    \label{fig:MoGconstraints}
\end{figure}

\subsection{Modified gravity}

Local Group dynamics in the context of modified gravity has already been discussed in the literature~\citep{McLeod:2019cfg,Benisty:2023ofi}. We are also able to constrain modified gravity theories. We only consider theories that are able to simultaneously drive the acceleration of the cosmic expansion (dark energy) while remaining compatible with solar system tests of gravity~\citep{Brax:2008hh,Benisty:2022lox}. For scalar tensor theories of gravity with conformal coupling to matter (via ${g}_{\mu\nu}^{(J)} =  \left(1+{2\beta}\frac{\phi}{m_{\rm Pl}}\right) g_{\mu\nu}^{(E)}$, where ${g}_{\mu\nu}^{\left(J\right)}$ is the metric in Jordan frame and $g_{\mu\nu}^{\left(E\right)} $ is the metric in the Einstein frame) gives an effective Newtonian constant: $G_{\text{eff}}/G_{\rm N} = 1 + \beta^2$ where $\beta$ is conformal strength~\citep{Brax:2019tcy}. By using the measured Planck value for $\Lambda$ and the Local Group mass, we can constraint any other modified gravity parameter. Fig~\ref{fig:MoGconstraints} shows the posterior distribution for $\Delta G_{\rm N}/G_{\rm N}$ that reads: 
\begin{equation}
 {\Delta G_{\rm N}/G_{\rm N} = 0.264 \pm 0.454. }
\end{equation}
Although the constraint on $G_{\rm N}$ is not as strong as other bounds, it is nonetheless interesting as it is probing deviations from Newtonian gravity on scales of Mpc. However, we note that in theories with screening we would expect the constraint on $G_{\rm N}$ to depend on the environment, being more screened in dense environments such as the solar system, see~\citep{Benisty:2022lox} and references therein. The Cassini bound $\Delta G_{\rm N}/G_{\rm N} = \left(2.1 \pm 2.5 \right) \times 10^{-5}$~\citep{Bertotti:2003rm} is on Solar system scales. Recently, \citet{Desmond:2020nde} have  claimed a five percent measurement of the Gravitational constant using Cepheid variable stars in detached eclipsing binaries in the Large Magellanic Cloud: $\Delta G_{\rm N}/G_{\rm N} = 0.07^{+0.05}_{-0.04}$. This result is interesting as it provides a measurement of $G_{\rm N}$ in another galaxy, but it is again probing scales of astronomical units rather than Megaparsecs.

\section{Discussion}  

The Cosmological Constant has a small but measurable effect on the motion of the Milky Way and Andromeda galaxies, which are the two largest members of the Local Group. This {\it Letter} shows that the ratio between the Keplerian period and $T_\Lambda = 2 \pi /(c \sqrt{\Lambda}) \approx 63.2 \,\, {\rm Gyr}$ controls the importance of the Cosmological Constant in the two body problem. Since the period of the Milky Way and Andromeda galaxies is about $20 \, {\rm Gyr}$ the Cosmological Constant is a roughly $10\%$ effect on the motion of the galaxies. In this limit, we provide an analytical solution for the two-body problem.

We show that a larger value of the Cosmological Constant gives a larger value for the mass of the Local Group. Based on the observationally derived measurements of the Local Group mass, we give a constraint on the Cosmological Constant only from dynamics.  {We find that upper limit on the Cosmological Constant is about $\sim 5.44$ times larger then the Planck value. Anticipating future gains in accuracy from {\it JWST} astrometry, we find that the upper bound could reduce to a direct detection of the $\Lambda$ with the value: $\left(1.67 \pm 0.8 \right) \Lambda_{PL}$ in the near future. This detection will be on very different scales than that probed by the CMB.}

The Local Group is also a new arena in which to test modified theories of gravity. We obtain new bounds on scalar-tensor theories. Although these are not as strong as Solar system constraints, they are still the first constraints on unprecedentedly large Megaparsec scales. Of-course the constraint could also be density dependent as in many popular models~\citep{Brax:2008hh} for example. Based on the condition for the domination of the Cosmological Constant in a binary system, we suggest for future research bounds on the Cosmological Constant and alternative gravity theories derived from binary galaxies in the Local Universe.

\begin{acknowledgments}
We thank Eugene Vasiliev for useful comments and code lines and to Jenny Wagner and Noam Libeskind for useful discussions.  {We thank for the anonymous referee for comments.}  DB gratefully acknowledges the supports of the Blavatnik and the Rothschild fellowships, as well as a Postdoctoral Research Associateship at the Queens' College, University of Cambridge. DB has received partial support from European COST actions CA15117 and CA18108 and the research grants KP-06-N58/5.
\end{acknowledgments}


\bibliography{sample631}{}
\bibliographystyle{aasjournal}



\end{document}